# Energy dependence of *W* values for protons in hydrogen

G.A.Korolev,* G.D.Alkhazov, A.V.Dobrovolsky, A.V.Khanzadeev, A.A.Vorobyov

*Petersburg Nuclear Physics Institute of National Research Centre "Kurchatov Institute",*

*188300, Gatchina, Russia*

**Abstract**. The mean energy *W* required to produce an ion pair in molecular hydrogen has been obtained for protons in the energy range between 1 MeV and 4.5 MeV. The *W* values were derived from the existing experimental data on elastic $\pi^- p$ scattering at the beam energy of 40 GeV. In the experiment, the ionization chamber IKAR filled with hydrogen at a pressure of 10 at served simultaneously as a gas target and a detector for recoil protons. For selected events of elastic scattering, the ionization yield produced by recoil protons was measured in IKAR, while the energy was determined kinematically through the scattering angles of the incident particles measured with a system of multi-wire proportional chambers. The ionization produced by *α*-particles from *α*-sources of $^{234}$U deposited on the chamber electrodes was used for absolute normalization of the *W* values. The energy dependence of *W* for protons in $H_2$ shows an anomalous increase of *W* with increasing energy in the measured energy range. At the energy of 4.76 MeV, the ionization yield for alpha particles is by 2% larger than that for protons.

## 1. Introduction

In gas-filled ionization detectors, the energy *T* of incident charged particles is determined via the mean number *N* of the ion pairs produced when the particles are completely stopped in the gas, provided the mean energy *W* required to produce an ion pair is known. The *W* value is defined as the ratio *T*/*N*. In the case of high-energy particles crossing a thin gas cell, one should know the differential value of the mean energy expended for formation of an ion pair, $w = dT/dN$, which is related to *W* as $w^{-1} = d/dT\,(T/W)$. The knowledge of the variation of *W* with the energy *T* is crucial in applications of ionization detectors in nuclear spectrometry and radiation dosimetry, as well as in radiotherapy. In the past few decades, the energy dependence of *W* values in various gases and gas mixtures was investigated in many laboratories. A good summary of theoretical and experimental data on the *W* values was given by the International Commission on Radiation Units and Measurements in the ICRU Report 31 [1]. This report also provides the suggested *W* values for the current usage. A review of recent developments in theory and experimental results on variation of *W* with energy can be found in [2].

* corresponding author: guerman@pnpi.spb.ru

The incident particle loses energy in its passage through a gas mainly by inelastic collisions with the gas molecules, resulting in their ionization and excitation. At high energies, the *W* value has little dependence on energy and generally approaches a constant value. On the other hand, a significant energy dependence of *W* for ionizing particles was observed in experiments with low energy particles. The existing data show an increasing *W* value with decreasing energy since the ratio of the ionization cross-section to those for non-ionizing processes at low energy becomes smaller. It should be noted that considerable discrepancies often exist for *W* values obtained in different experiments because of difficulties to measure *W* with high precision.

The energy dependence of *W* for protons and heavier particles was found to be different from that for electrons. While the *W* values for electrons typically decrease smoothly with increasing energy, theoretical and experimental *W* values for heavy particles exhibit an extraordinary energy dependence that appears at certain energies [1–3]. In particular, in the case of protons a broad minimum was observed in some gases ($N_2$, $CO_2$, air) at the energy of about 20 keV [2]. Rather scarce information exists on *W* values for protons in hydrogen, the gas frequently used in the proton recoil measurements. The *W* values measured in the energy range 1 keV $\leq T \leq$ 100 keV [4] show a steep decrease with increasing energy up to about 5 keV, followed by a flat minimum around 40 keV. No experimental data exist at higher energies apart from the *w* value measured at 340 MeV [5].

In the present work, the *W* values for recoil protons in hydrogen in the energy range from 1 MeV to 4.5 MeV are derived from the raw experimental data on small-angle elastic $\pi^- p$ scattering at the beam energy of 40 GeV measured with the ionization recoil detector IKAR at IHEP [6]. To check the validity of the used analysis, the energy dependence of *W* for recoil α-particles was also derived from the data on elastic $\pi^{-4}$He scattering measured in the experiment with the same beam [7] when IKAR was filled with the mixture of $^4$He + 11% $H_2$.

## 2. Method of measurement

The detector IKAR used in the discussed experiments was designed and built at PNPI [8, 9]. IKAR is a special ionization chamber which serves simultaneously as a gas target and a recoil particle detector with a high energy resolution. IKAR was used in several experiments on small-angle elastic scattering of high-energy hadrons and light ions at PNPI [10], IHEP [6, 7], CERN [11], Saclay [12], and GSI [13].

The aim of all these experiments was to measure the absolute differential cross sections d$\sigma$/d$t$ for elastic scattering of the beam particles on protons, deuterons and $^4$He nuclei. In order to measure d$\sigma$/d$t$ with high precision, it is essential to have a precise absolute calibration of the $t$-scale. The four momentum transfer squared $t$ was determined by the kinetic energy $T_R$ of the recoil particle

$$|t| = 2mT_R, \qquad (1)$$

$m$ being the mass of the recoil particle. The value of $T_R$ was obtained by measuring the charge of electrons released in the process of ionization of the gas in IKAR. Therefore, one should know how the ionization yield is related to the energy $T_R$. As a first approximation, the calibration of the energy scale was made with the aid of $\alpha$-sources deposited on cathodes. A kinematical relation between the energy of the recoil particle and the scattering angle $\theta$ of the projectile was used for a more precise absolute $t$-scale calibration. At small scattering angles $\theta$, the $t$-value is related to $\theta$ as

$$|t| = p^2 \theta^2, \qquad (2)$$

were $p$ is the momentum of the beam particle. The scattering angle $\theta$ was measured with multi-wire proportional chambers placed ahead and behind the spectrometer IKAR. Combining Eq. (1) and Eq. (2), we obtain the recoil energy as

$$T_R = p^2\theta^2/2m. \qquad (3)$$

Our analysis has shown that the determined energy $T_R$ in the studied range of $T_R$ is a linear function of the ionization yield (the anode amplitude $V_R$) produced by the recoil particles:

$$T_R = kV_R + T_0, \qquad (4)$$

where $k$ and $T_0$ are the parameters which depend on the type of the gas, pressure and electric field [12]. The value of $k$ also depends on the amplifier gain. Obviously, such a calibration contains information on the energy behaviour of the $W$ values which may be obtained as a by-product in the experiment.

## 3. Experimental set-up and procedure

We analysed the data on elastic $\pi^- p$ scattering [6] at the beam energy of 40 GeV to derive the $W$ values for recoil protons in hydrogen. In the experiments at IHEP [6, 7], both the scattering angle $\theta$ of the scattered pions and the yield of ionization produced by recoil particles were measured. The main component of the set-up was the recoil detector IKAR [9, 11]. IKAR is a pulse ionization chamber operating with electron collection. The chamber was filled with pure hydrogen at a pressure of 10 at. IKAR consists of six independent identical modules. Each module is an axial ionization chamber, which contains an anode plate, a cathode plate, and a grid, all the electrodes being arranged perpendicular to the beam direction. For events of elastic scattering at small angles, the recoil proton track is almost parallel to the chamber electrodes. The amplitude and time analyses of signals from the electrodes of IKAR provide the recoil energy $T_R$, or d$E$/d$x$ for the recoil particles which leave the active volume, the coordinate $Z_R$ (along the chamber axis) of the interaction vertex in the cathode − grid space, and the scattering polar angle $\theta_R$ of the recoil proton.

The energy resolution of IKAR was 60 keV (FWHM). In order to have an accurate energy calibration for the amplitudes $V_R$ of the anode pulses, complete collection of all electrons formed by recoil protons in the ionization process is necessary. The presence of electronegative gas impurities in the ionization chamber leads to some losses of electrons due to adhesion. Furthermore, the grid in the chamber is not an ideal shield against the induction effect of positive ions. Therefore, the signal from the anode may be reduced depending on the position of the track in the cathode − grid space [8]. A special procedure was developed to introduce the necessary correction. For this purpose, $\alpha$-sources of $^{234}$U were deposited on all the cathodes and grids. The difference $\Delta V_\alpha$ in the positions of two measured $\alpha$-peaks corresponding to $\alpha$-particles emitted from the grid and cathode sources was used to control the purity of the gas in the chamber [11]. The correction $\delta$ to the amplitude $V_R$ as a function of $Z_R$ was applied, which took into account both the loss of electrons through adhesion and the limited transparency of the grid:

$$\delta(Z_R) = \frac{(d - Z_R)}{d} \frac{\Delta V_\alpha}{V_{\alpha c}}. \tag{5}$$

Here, $d$ is the cathode−grid distance, $V_{\alpha c}$ is the mean amplitude of the pulses produced by the $\alpha$-particles from the cathode $\alpha$-source. The value of $\Delta V_\alpha / V_{\alpha c}$ was about 1%.

The correction on the charge recombination may be important for the chamber operating at high pressure. In a special study with the ionization chamber, the saturation characteristics of the pulse height $V_\alpha$ for α-particles against the voltage $U$ on the cathode at various pressures of hydrogen were measured. To separate recombination effects from electron adhesion only the signals obtained for particles emitted from the grid α-sourse were used. The charge collection efficiency $f_\alpha$ as a function of inverse of the electric field strength $d/U$ at various pressures is shown in Fig. 1. The linear extrapolation of the observed plot to $d/U = 0$ furnishes the collection efficiency corresponding to complete saturation. For the operating conditions of the experiment [6] (electric field strength 1.5 kV/cm, pressure 10 at), the relative charge loss for α-particles due to recombination was found to be 1.8% with the respective collection efficiency $f_\alpha = 0.982$. Since the ionization density for protons is much smaller than that for alphas, the recombination effect for protons under the used conditions was assumed to be negligible, $f_p \approx 1$. The charge recombination of α-particles in the mixture of He + 11% H$_2$ under the operating conditions of the experiment [7] was found to be 0.3%.

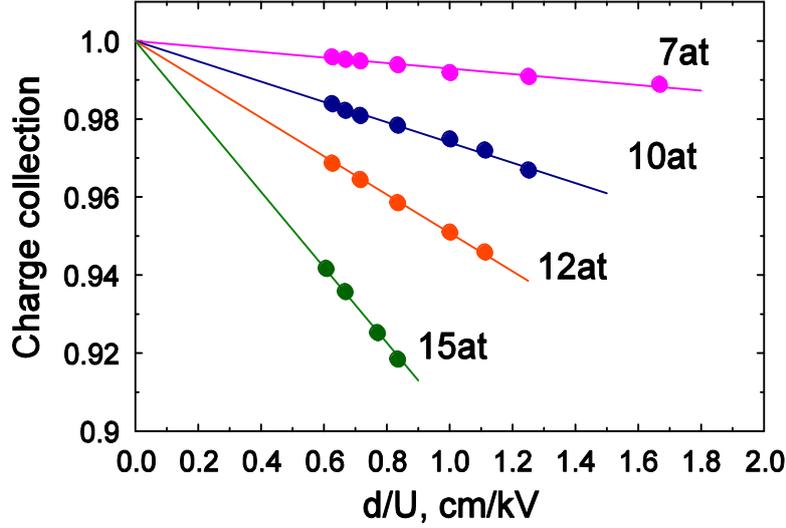

Fig. 1. Recombination for α-particles in H$_2$ at various pressures. The charge collection efficiency versus inverse of the electric field strength $d/U$. Cathode – grid distance is 100 mm.

An accurate determination of the beam momentum $p$ was performed in the experiment on elastic $\pi^-$ $^4$He scattering [7]. The $T_R$ scale was calibrated using α-particles of kinetic energy $E_\alpha$ from the decay of $^{234}$U. The recoiling particles in this case were also α-particles. Therefore, the α-source signals for events with $T_R^* = E_\alpha$ represent here an absolute calibration reference, which does not depend on any assumptions about the relation between the ionization charge and the recoil energy. In practice, the value of $\theta^*$ was deduced from the correlation between $V_R$ and $\theta$ in the measured data as the mean scattering angle corresponding to $V_R^* = V_\alpha$. The resolution of the scattering angle $\theta$ measured with ordinary multi-wire proportional chambers was $\sigma_\theta = 0.17$ mrad. To this end, the value of the beam momentum $p$ was determined as $p = \sqrt{2m_\alpha E_\alpha} / \theta^*$ with the precision of 0.2%. Details of this method are described in Ref. [14].

## 4. Results

In order to investigate the energy dependence of $W$ for protons in $H_2$, the experimental data [6] were analysed in the following way. For selected events of elastic scattering, the data were binned into small intervals of $V_R$. The values of the anode amplitudes $V_R$ were corrected for adhesion and the grid inefficiency (5), the mean scattering angle $\theta$ was found for each interval, and the corresponding value of $T_R$ was calculated using Eq. (3). To relate calibration (4) to the energy of the $\alpha$-line, Eq. (4) was rewritten in the following form

$$T_R = \frac{(E_\alpha - T_0)}{\beta V_\alpha} V_R + T_0 \qquad (6)$$

and fitted to the data, the values of $T_0$ and $\beta$ being considered as free parameters. Here $\beta = V_R^* / V_\alpha$ is the ratio of the ionization yields produced by protons and alphas of the same energy $T_R^* = E_\alpha$. The value $E_\alpha = E_\alpha^{(0)} - \Delta E_{abs}$, where $E_\alpha^{(0)} = 4.7746(14)$ MeV [15], and $\Delta E_{abs} = 10$ keV is the correction on the energy absorption in the matter of the $\alpha$-source [11, 16]. As a result of the fit, we have obtained the parameters $\beta = 0.998 \pm 0.005$ and $T_0 = (-62.1 \pm 6.6)$ keV for the operation conditions given above. The experimental relation between $V_R$ and $T_R$ and the fitted straight line are shown in Fig. 2. The experimental data on the ratio of $T_R/V_R$ were used to obtain the relative energy dependence of $W$ for protons in the energy range $1.0 \leq T_R \leq 4.6$ MeV. A similar procedure was applied to the selected elastic $\pi^-$ $^4$He-scattering events [7]. But in addition, a correction for recombination was applied. The ionization density is maximal at the end of the track, so in the first approximation the charge loss was assumed to occur only there. Having in mind that the measured recombination $\Delta V$ at the energy $T_R^* = E_\alpha$ is 0.3% of the value $V_R^* = V_\alpha$, all the amplitudes $V_R$ were increased by the value $\Delta V = 0.003 V_\alpha$. A straight line fit (6) describes the data well (see Fig. 3) in the energy interval $1.7 \leq T_R \leq 10.3$ MeV with $T_0 = (19.2 \pm 6.3)$ keV and $\beta$ fixed to $\beta = 1$. The relative energy dependence of $W$ for $\alpha$-particles in the $^4$He – $H_2$ mixture was derived from the measured $V_R - T_R$ correlation. The absolute normalization of the $W$ values was obtained using the results on measuring the ionization produced by $\alpha$-particles of $^{239}$Pu ($E_\alpha = 5.14$ MeV) in the He – $H_2$ mixture [17]. The $W$ value measured in [17] for the $^4$He + 11% $H_2$ mixture is $W = (31.9 \pm 0.3)$ eV. The variation of $W$ with the energy normalized in such a way is shown in Fig. 4 and is presented in the Appendix.

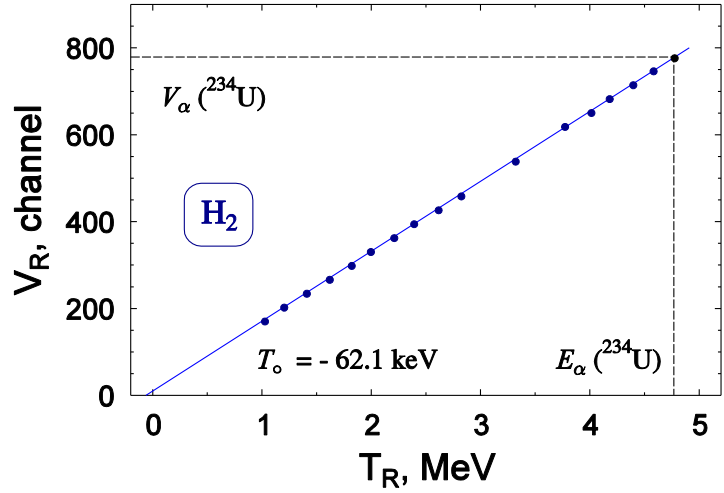

Fig. 2. Experimental relation (dots) between the anode amplitude $V_R$ and the proton recoil energy $T_R$ as measured in the elastic $\pi^- p$ scattering [6] and the straight line fitted with Eq. (6). The recoil energy $T_R$ was calculated from the measured scattering angle $\theta$ of the pions.

The relative energy dependence of $W_p$ for protons in $H_2$ was normalized by using as a reference the known value $W_\alpha$ for $\alpha$-particles. Under the condition $T_R^* = E_\alpha$, a relation between the ionization yield produced by protons and by $\alpha$-particles of the same energy taking into account the charge recombination reads as $f_p T_R^* / W_p = \beta\, f_\alpha E_\alpha / W_\alpha$. Substituting all the measured values, we obtain $W_p = 1.02\, W_\alpha$. According to the recommendation of the ICRU [1], $W_\alpha = (36.43 \pm 0.63)$ eV at $E_\alpha = 5.3$ MeV. We admitted that $W_\alpha$ is the same at $E_\alpha = 4.76$ MeV and performed absolute normalization of the $W_p$ values. The values $W_p$ obtained in the present work are shown in Fig. 5 together with the values determined in [4, 5]. The $W_p$ values are also given in the Appendix. Note, that the result of Ref. [5] for 340 MeV is replaced by the value $w = 36.3$ eV revised in [4].

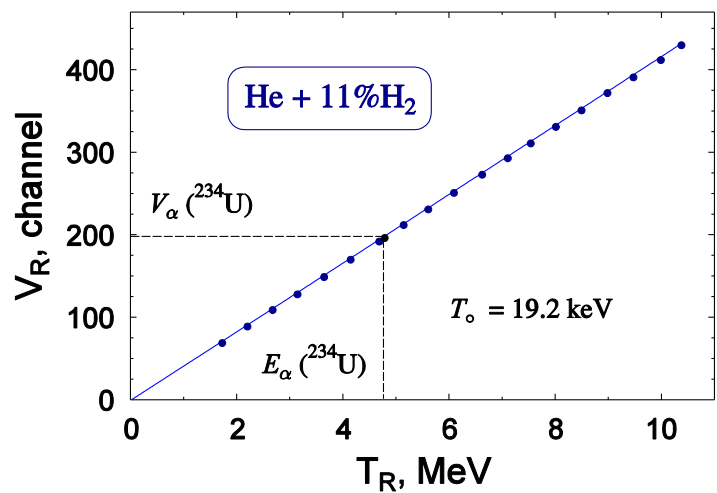

Fig. 3. The same as in Fig. 2, but for the recoil $\alpha$-particles in the elastic $\pi^{-4}$He scattering [7]. The anode amplitudes $V_R$ are corrected for recombination.

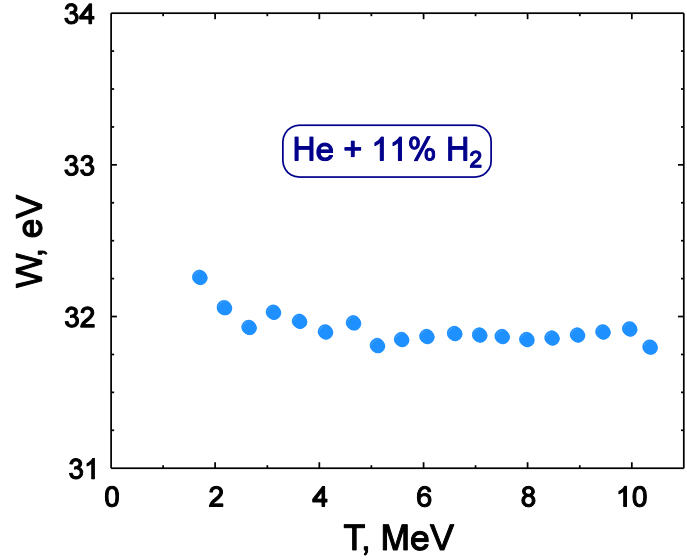

Fig. 4. Energy dependence of the $W$ value for $\alpha$-particles in the He + 11% $H_2$ gas mixture.

5. Discussion

We have obtained the typical monotonic decrease of the $W$ values for alphas in the He – $H_2$ mixture in the MeV energy range. In contrast to that, our results for protons in $H_2$ obtained with the same method show an increasing $W$ value by about 5% in the energy range from 1 MeV to 3 MeV. At higher energy, $W$ remains constant. One can see in Fig. 3 that the data [4] obtained at low energy up to 100 keV can be smoothly extended to the data of the present investigation. In comparison with the minimum at about 40 keV found in [4], our value of $W$ at 4 MeV is larger by more than 17%. Our overall relative uncertainty in $W$ is estimated to be not larger than 2%, whereas the $W$-error quoted in Ref. [4] is smaller than 1%.

The data obtained in other experiments with IKAR do not permit to find accurate absolute $W$ values for recoil particles due to lack of all the necessary raw experimental information.

Nevertheless, in the case of recoil protons in $H_2$, the same dependence of the relative ionization yield on the proton energy as discussed above was obtained in the experiments at CERN [11] and PNPI [10]. In the energy calibration of amplitudes $V_R$ according to Eq. (4), the value of the parameter $T_0$ was also found to be negative. This means that the $W$ value increases with the energy in the MeV energy range. On the other hand, positive values of $T_0$ were determined in the analysis of the energy dependence of the ionization yield produced by recoil $\alpha$-particles in the $He-H_2$ gas mixture [18, 19] and by recoil protons in $CH_4$ [12], which correspond to smoothly decreasing $W$ with energy. This conclusion agrees well with the general trend for a monotonic decrease of $W$ with the energy increasing [2], thus giving confidence to the method used in the performed analysis.

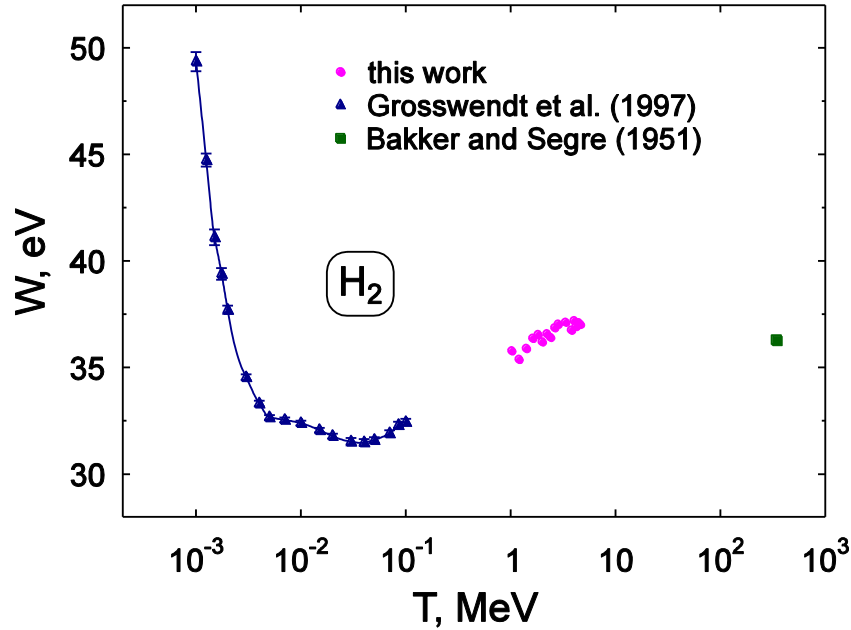

Fig. 5. Experimental $W$ values for protons in $H_2$ obtained in this work together with the results of $W$ measurements [4] at lower energies. The differential $w$-value measured for 340 MeV protons [5] and revised in Ref. [4] is 36.3 eV. This $w$ value may be taken as the high-energy $W$ value.

A certain structure in the energy behaviour of $W$ values for protons was observed in some gases other than $H_2$, though marked deviations exist between the data sets reported by different authors. A minimum in the $W$ values was found in $N_2$, $CO_2$ [20–22] and air [20] at energies near 20 keV. Note that the most pronounced *W* structure is reported for $H_2$ in the present work.

The energy dependence of the *W* values for protons in gases is not yet satisfactorily explained. The experimental data indicate that *W* values in some gases do not decrease monotonically. To explain the formation of the minimum in the energy dependence of *W* for protons in these gases, it was assumed that a high cross section for the charge exchange processes with small energy losses for hydrogen projectiles at the energy below 50 keV may be responsible for the enlarged ionization yield [20–23]. An analytical model based on the continuous slowing-down approximation was applied to calculate the *w* and *W* values for protons [3, 4, 23]. In that model, the experimental data for the stopping power, ionization and charge exchange cross sections for hydrogen ions and atoms, and the contributions of secondary electrons were taken into account in the determination of the ionization yield. Rather satisfactory agreement between the experimental data and the calculated values was obtained for $N_2$, $CO_2$ and Ar. It was shown that the charge exchange process leads to production of additional ion pairs, resulting in the reduced *W* values. However, the model does not describe the *W* behavior in $H_2$ [4]. Perhaps, the reason of this is in the considerable collateral ionization originated from excited hydrogen atoms in dissociative processes with formation of $H_3^+$ ions [4, 24].

**Conclusion**

The experimental values of the mean energy *W* expended to produce an ion pair in $H_2$ have been obtained for protons in the MeV range. Our data show an anomalous increase of *W* values with increasing energy and are consistent with the results of Ref. [4] at lower energies, thus demonstrating an oscillatory behaviour of the *W* values. Experimental results for some other gases reported by different authors also show structured energy dependence, but the most pronounced minimum is observed in $H_2$. Unfortunately, an adequate theory has not yet been developed to explain the extraordinary energy behaviour of *W* for protons in hydrogen. A competition between ionization (including charge-exchange effects) and excitation processes may be responsible for such a structured energy dependence of *W*. Our measurements also have shown that the ionization yield for *α*-particles in $H_2$ is by 2% larger than that for protons at the energy of about 5 MeV.

This work was supported by the grant of the President of the Russian Federation # NSh – 393.2012.2

**Appendix**

Table 1.  Experimental *W* values for protons stopped in hydrogen.

| *T*, MeV | *W*, eV | *T*, MeV | *W*, eV |
|----------|---------|----------|---------|
| 1.016 | 35.80 | 2.605 | 36.88 |
| 1.192 | 35.40 | 2.813 | 37.05 |
| 1.399 | 35.91 | 3.310 | 37.14 |
| 1.609 | 36.39 | 3.762 | 36.77 |
| 1.810 | 36.56 | 4.003 | 37.21 |
| 1.985 | 36.22 | 4.170 | 36.94 |
| 2.199 | 36.60 | 4.386 | 37.12 |
| 2.381 | 36.43 | 4.572 | 37.04 |

Table 2.  Experimental *W* values for *α*-particles in the He+11%$H_2$ gas mixture.

| *T*, MeV | *W*, eV | *T*, MeV | *W*, eV |
|----------|---------|----------|---------|
| 1.698 | 32.26 | 6.592 | 31.89 |
| 2.171 | 32.06 | 7.073 | 31.88 |
| 2.645 | 31.93 | 7.506 | 31.87 |
| 3.113 | 32.03 | 7.979 | 31.85 |
| 3.614 | 31.97 | 8.462 | 31.86 |
| 4.112 | 31.90 | 8.951 | 31.88 |
| 4.651 | 31.96 | 9.440 | 31.90 |
| 5.110 | 31.81 | 9.955 | 31.92 |
| 5.575 | 31.85 | 10.345 | 31.80 |
| 6.059 | 31.87 | | |